\documentstyle[aaspp4,psfig]{article}
\newcommand{\phsys}{${\rm A}^{1}\Pi_{u}-{\rm X}^{1}\Sigma_{g}^{+}$}
\newcommand{\cosys}{${\rm X}^{1}\Sigma^{+}$}

\begin{document}
\title{The circumstellar shell of the post-AGB star HD\,56126: 
the $^{12}$C$^{12}$C/$^{12}$C$^{13}$C isotope ratio
and $^{12}$C$^{16}$O column density}
\author{Eric J. Bakker\footnote{
present address:
TNO-FEL, Electro Optics Group, 
P.O. Box 96864, 2509 JG, The Hague, The Netherlands,
e.j.bakker@fel.tno.nl}  
and David L. Lambert}
\affil{Department of Astronomy and  W. J. McDonald Observatory, 
University of Texas, Austin, TX\,78712-1083, USA}
\affil{ebakker@astro.as.utexas.edu, dll@astro.as.utexas.edu}

\begin{abstract}
We have made the first detection of circumstellar absorption
lines of the $^{12}$C$^{13}$C \phsys~ (Phillips) system  1-0 band
 and     the $^{12}$C$^{16}$O \cosys~ first-overtone     2-0 band in 
the spectrum of the post-AGB star HD\,56126
(IRAS\,07134+1005). 
All current detections
of circumstellar molecular absorption lines towards HD\,56126
($^{12}$C$_{2}$, $^{12}$C$^{13}$C, $^{12}$C$^{14}$N, $^{13}$C$^{14}$N, 
and $^{12}$C$^{16}$O) yield the same 
heliocentric velocity of $v_{\rm CSE}=77.6\pm0.4$~km~s$^{-1}$.
The $^{12}$C$_{2}$, $^{12}$C$^{13}$C, and $^{12}$C$^{16}$O lines give
rotational temperatures and integrated column densities of 
$T_{\rm rot} = 328\pm37$~K, $\log N_{\rm int}=15.34\pm0.10$~cm$^{-2}$,
$T_{\rm rot} = 256\pm30$~K, $\log N_{\rm int}=13.79\pm0.12$~cm$^{-2}$, and
$T_{\rm rot} =  51\pm37$~K, $\log N_{\rm int}=18.12\pm0.13$~cm$^{-2}$
respectively.
The rotational temperatures are lower for molecules with a higher permanent
dipole moment. 
Derived relative column densities ratios are 
$^{12}$C$_{2}$/$^{12}$C$^{13}$C$=36\pm13$, and
$^{12}$C$^{16}$O/($^{12}$C$_{2}$+$^{12}$C$^{13}$C)$=606\pm230$.
Combined with data from Paper~III we find
relative column densities of $^{12}$C$^{16}$O/($^{12}$C$^{14}$N+$^{13}$C$^{14}$N)$=475\pm175$ and
$^{12}$C$^{14}$N/$^{13}$C$^{14}$N$=38\pm2$.

Under chemical equilibrium
conditions, $^{12}$C$^{13}$C is formed twice as easily as $^{12}$C$_{2}$.
The isotopic exchange reaction for $^{12}$C$_{2}$  
is too slow to significantly alter the $^{12}$C$_{2}$/$^{12}$C$^{13}$C ratio
and the $^{12}$C$_{2}$ to $^{12}$C$^{13}$C ratio
a good measure of half the carbon isotope ratio: 
$^{12}$C/$^{13}$C=$2\times$$^{12}$C$_{2}$/$^{12}$C$^{13}$C=$72\pm26$. 
This is in 
agreeement with our prediction that the isotopic exchange
reaction for $^{12}$C$^{14}$N  is efficient
and our observation in Paper~III of $^{12}$C$^{14}$N/$^{13}$C$^{14}$N$=38\pm2$.

A fit of the 
C$_{2}$ excitation model of van Dishoeck \& Black 
(\cite{vandishoeckblackmodel})
to the relative population distribution of C$_{2}$ yields
$n_{\rm c} \sigma / I=3.3 \pm 1.0 \times 10^{-14}$. At $r\simeq 10^{16}$~cm
this translates in $n_{\rm c}=1.7 \times 10^{7}$~cm$^{-3}$ and
$\dot M \simeq 2.5 \times 10^{-4}$~M$_{\odot}$~yr$^{-1}$.

\noindent
\keywords{line: identification       -- 
          molecular data             -- 
          molecular processes        --
          stars: AGB and post-AGB    -- 
          stars: circumstellar matter--
          individual stars: HD\,56126 
}
\end{abstract}

\section{Introduction}

HD\,56126  (IRAS\,07134+1005) is in the post-AGB stage (also referred to as 
the Pre-Planetary Nebulae stage) of stellar evolution.
This is the relative short transition stage from the Asymptotic
Giant Branch (AGB) to the White Dwarf (WD) phase
(Iben~\cite{iben}). Low-mass
stars (0.8 M$_{\odot} \leq$ M$_{\ast} \leq$ 8.0 M$_{\odot}$)
may experience up to three dredge-up episodes. The first occurs 
on the Red Giant
Branch (RGB). The second on the Early-AGB (E-AGB) is limited
to stars with initial masses in the range of 4 to 8~M$_{\odot}$.
The third occurs on the
Thermal Pulsating-AGB (TP-AGB). The last dredge-up changes the
surface abundances of the star most drastically: most noticeable the carbon 
and s-process elements abundances enhanced
(Forestini \& Charbonnel~\cite{forestinicharbonnel}). Several studies of
the photospheric abundance of post-AGB stars have indeed shown
that some of these stars are carbon rich and have enhanced abundances
of s-process elements 
(Luck \& Bond~\cite{luckbond};  
 Klochkova~\cite{klochkova};    
 van Winckel~\cite{vanwinckel}; 
 Reddy et al.~\cite{reddyetal},\cite{reddybakkerhrivnak}; 
 Decin et al.~\cite{denin}). 
Since AGB stars lose mass through a dense stellar
wind, post-AGB stars are surrounded by a shell of material which
expands at  typically 3 to 30~km~s$^{-1}$.
This material constitutes 
the circumstellar environment (CSE) of the star. The CSE material mixes
with interstellar clouds and forms the parent material for
the next generation of stars. In this manner, nucleosynthesis products
are used to build  new  stars and the Galaxy
is enriched with heavy elements (elements more massive than helium).
The study of the chemical structure of the CSE of evolved stars is
therefore of importance to understand the chemical enrichment of
the Universe.

Stellar 
evolution theory needs constraints from observations.
The  $^{12}$C/$^{13}$C suits this goal very well since during the 
AGB evolution $^{12}$C is formed in the helium burning layer, while
$^{13}$C is formed as a by-product in the hydrogen burning layer. The
observed ratio of these two isotopes depends on the efficiency
and the relative strength of these two layers,
and on the efficiency of the dredge-up processes.

Theoretical models of the chemical structure of the CSE of 
highly evolved stars have mainly concentrated on understanding
the well-studied, and well-observed carbon star IRC\,+10216
(for references see Olofsson \cite{olofsson}). When
a star reaches the tip of the AGB, the mass-loss rate has its
maximum strength and drops several orders of magnitude after
the star leaves the AGB. Material
ejected during the AGB phase slowly expands from the star and leaves
a cavity in the CSE which will grow with time. Eventually leading to
a cavity the size of a planetary nebulae. 
The elemental composition of the CSE is fixed after the material leaves
the surface of the star. In contrast, 
the molecular composition of the
CSE is constantly changing under the influence of the stellar
and interstellar radiation fields: molecules are destroyed and
formed continuously. Most studies of the 
CSE make use of molecular line emission in the sub-millimeter and
radio (e.g. CO and HCN) while recently the ISO 
satellite has allowed a study of molecular
lines in the infrared.
Emission lines of spatially unresolved
sources provide no positional information, and only partial
velocity information (the radial component).
This is obviously not the 
case for absorption lines since they are formed in a pencil-beam
towards the continuum source and probe therefore a 
well defined region of the CSE. Secondly, a molecular 
absorption line spectrum
has many lines allowing a study of the population of
the energy levels of the molecules and  processes like
optical pumping. Absorption lines 
therefore yield information additional to that already available
from emission line studies.
As part of an ongoing effort to study the characteristics of post-AGB
stars, we have been studying circumstellar $^{12}$C$^{14}$N 
and $^{12}$C$_{2}$ lines in the 
optical spectra
of post-AGB stars (referred to as C2CN stars) which show the unidentified 
21~$\mu$m feature
(Kwok et al. \cite{kwok89},\cite{kwok95}; Hrivnak \& Kwok \cite{hrivnakkwok};
Justtanont et al. \cite{justtanont}). 
These electronic bands allow an accurate determination
of the expansion velocity of the CSE, rotational temperature,
column density, and molecular column density ratios
(Bakker et al. \cite{paperI} (Paper~I), \cite{paperII} (Paper~II), 
Bakker \& Lambert \cite{paperIII} (Paper~III)).

In Paper~III we presented measurements of the 
ratio CN/$^{13}$CN
(from now on C means $^{12}$C,
 N means $^{14}$N, and
 O means $^{16}$O). 
We found a ratio of $38\pm2$ but argued that 
the true C/$^{13}$C ratio
is likely closer to $67$ owing to the isotopic exchange reaction
($
{\rm ^{13}C^+} + {\rm      CN}   \rightleftharpoons 
{\rm      C^+} + {\rm ^{13}CN} + \Delta E $).
In this paper (IV) we present measurements
of the C$_{2}$/C$^{13}$C ratio.  
Combining these
two independent isotope ratios allows us to constrain
the isotope exchange reaction and give a more accurate determination
of the true C/$^{13}$C ratio in the 
CSE surrounding HD\,56126.

In Sec.~2 we discuss the optical and infrared
observations and the source of the 
data on the equivalent widths used in our analysis. 
Sec.~3 describes the  sources from which we obtained
the molecular data (C$_{2}$, C$^{13}$C, and CO).
Sec.~4 is the analysis and Sec.~5 the discussion.

\section{Observations and equivalent widths}

\subsection{Optical}

\begin{figure*} 
\centerline{\hbox{\psfig{figure=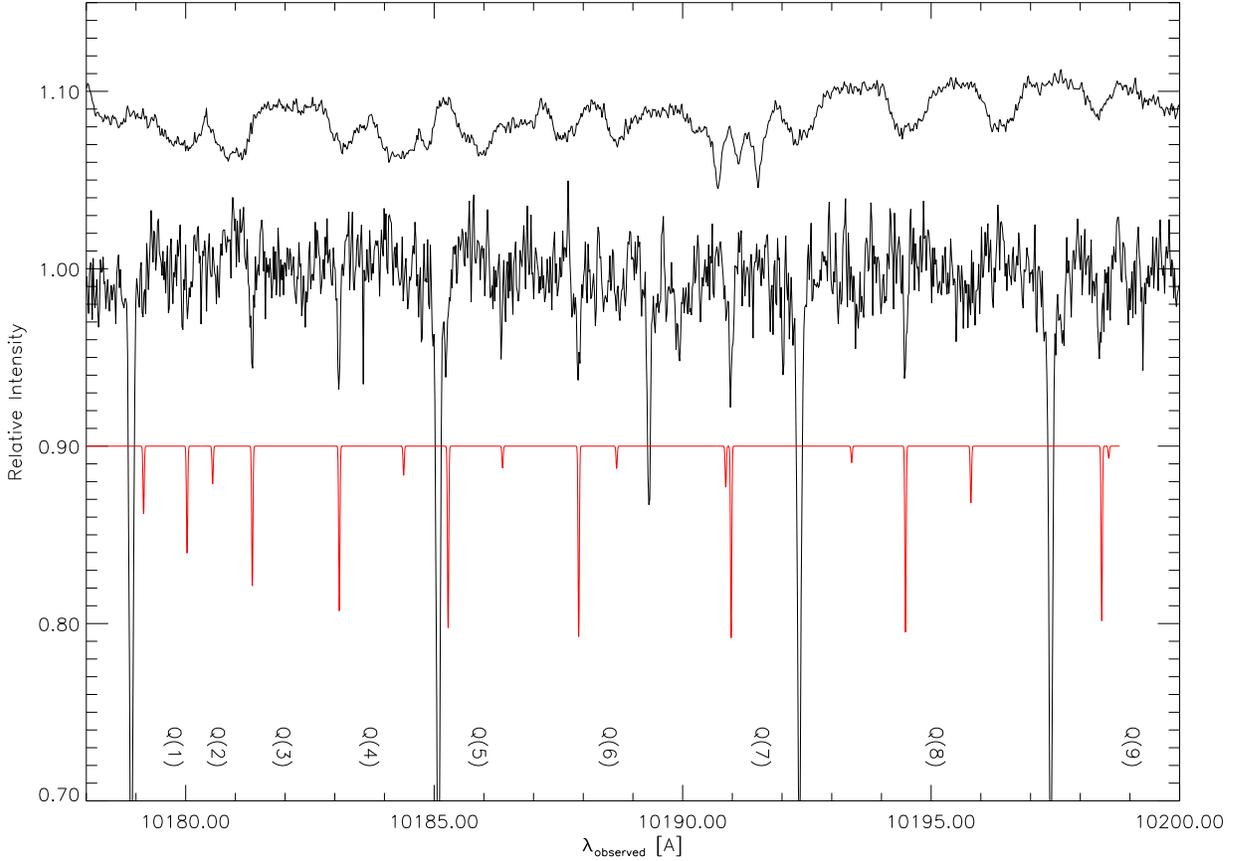,width=\textwidth,angle=90}}}
\caption{
C$^{13}$C Phillips system 1-0 band towards HD\,56126.
The spectrum of $\beta$ Ori
has been over plotted such that artifacts (telluric features, fringes etc.)
are aligned in both spectra.
The solid spectrum is the synthetic spectrum
($T_{\rm rot}=256$~K, $\log N_{\rm int}=13.79$~cm$^{-2}$, and $b=0.51$~km~s$^{-1}$) 
convolved to a spectral resolution of $R=130\,000$. The $f_{(1-0)}$-value
has been multiplied by 0.62 to obtain a better fit (see
Sec.~\ref{sec_bpar} for details).
The five strong lines which are not identified as C$^{13}$C are from
C$_{2}$ 1-0 (P(6), Q(12), R(22), P(8), and Q(14)).
For clarity,
only the Q-branch lines are marked, but the synthetic spectrum is based
on all allowed transitions.
}
\label{fig_specc2}
\end{figure*}

Spectra of the C$_{2}$ and
C$^{13}$C \phsys~ (Phillips) system  1-0 band have been
obtained using the coud\'{e} cross-dispersed echelle spectrograph 
(Tull et al. \cite{tull}) of the 2.7 meter
Harlan J. Smith telescope of the W.J. McDonald Observatory
(Table~\ref{tab_log}). During
three consecutive nights in January 1998, 
36 exposures of each 30 minutes on HD\,56126
have been made. Each exposure gives a CCD
frame  (TK3 with $2048\times2048$ pixels) with 18 orders,
with each order covering almost 19~\AA.
A log of the observations is presented in Table~\ref{tab_log}.
Each exposure is individually reduced
using the echelle package within IRAF: trimmed, scattered
light subtracted, flat fielded, orders extracted, 
and wavelength calibrated.
The 36 spectra are combined on an heliocentric wavelength scale,
 and finally
continuum corrected to obtain the co-added 
spectrum (Fig.~\ref{fig_specc2}). 
The analysis presented is based on this
final co-added spectrum. The spectral resolution, 
$R=\lambda/\delta \lambda=130\,000$ ($0.^{\prime\prime}6$ wide slit),
has been determined from the $FWHM$ of the C$_{2}$ lines
where the intrinsic width is expected to be less than the
instrumental width. No ThAr lines are available
in the order in which C$_{2}$ occurs, but 
ThAr lines available in other orders give  comparable numbers.
In order to be able to remove telluric lines 
and features that result from fringes on the CCD, we
observed the hot star $\beta$ Ori.

An interval 10160-10220 \AA~ was observed that includes the 
high excitation C$_{2}$ lines ($J^{\prime\prime}=4$ to 22)
and low-excitation C$^{13}$C lines ($J^{\prime\prime}=3$ to 9).
Equivalent widths (Table~\ref{tab_wcc}) 
were measured in various ways using the
tools available in IRAF/SPLOT.

\subsection{Infrared}

CO \cosys~ first-overtone
2-0 infrared spectra towards HD\,56126 were acquired with the 3.0 meter
NASA Infrared Telescope Facility (IRTF) on Mauna Kea, Hawaii, using 
the single-order cryogenic echelle spectrograph CSHELL
(Tokunaga et al. \cite{tokunagaetal}; Greene et al. \cite{greeneetal})
(Table~\ref{tab_log}).
Spectra were obtained with a $0.^{\prime\prime}5$ wide slit and  
$R=\sigma/\delta\sigma=43\,000$ (Greene \& Denault \cite{greenedenault}). 
The recorded spectrum covers about 10~cm$^{-1}$.
HD\,56126 was observed at two settings. One for
the 0,1,2 $J^{\prime\prime}$ levels,
and one the 4,5,6 $J^{\prime\prime}$ levels. Observations of 
$J^{\prime\prime}=3$ were not attempted since this line 
fell together with a strong telluric feature.
In order to remove the sky background, the telescope was nodded
$10^{\prime\prime}$ east-west between two successive exposures. The hot star
HR\,2763 ($\lambda$ Gem) was observed to act as a reference spectrum to remove
telluric features and instrumental artifacts.

The infrared data were reduced with IRAF.
First two successive frames were subtracted
to remove the bias and sky background. The result
was divided by the flat field minus dark frame. The spectra
were extracted and the wavenumber calibration was made on 
identified telluric lines using rest wavenumbers from 
the AFCRL absorption line catalog for basic input data
(ATMOS program, using software written by E.N. Grossman),
in combination with the atlas of Hinkle et al. (\cite{hinkleetal}).

\begin{figure*} 
\centerline{\hbox{\psfig{figure=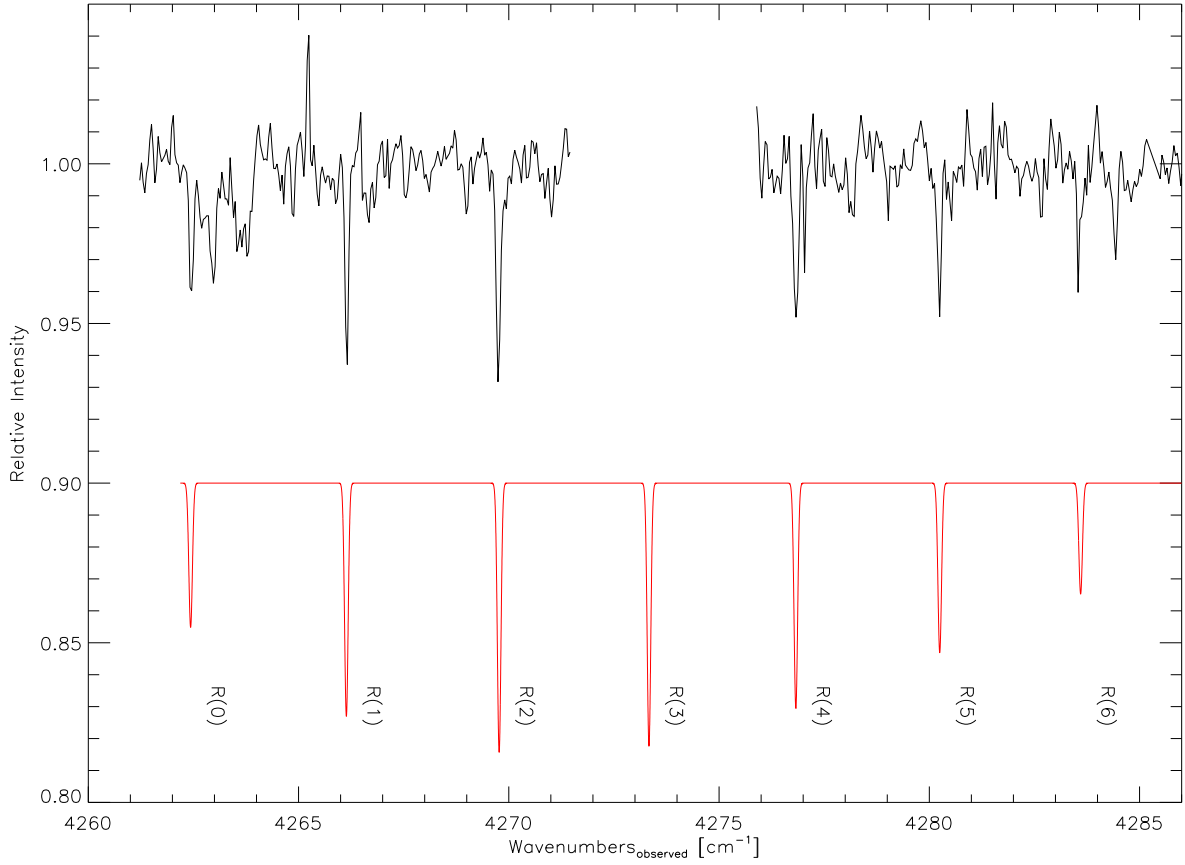,width=\textwidth,angle=90}}}
\caption{
CO first overtone band towards HD\,56126. The solid spectrum is the 
synthetic spectrum 
($T_{\rm rot}=51$~K, $\log N_{\rm int}=18.12$~cm$^{-2}$, and $b=0.51$~km~s$^{-1}$) 
convolved to a spectral resolution of R=$43\,000$.
The R-branch lines have been labeled. R(3) was not observed as
it was blended by a strong telluric line.
}
\label{fig_specco}
\end{figure*}

The resulting spectrum of HD\,56126 was divided by that of the 
hot star and continuum corrected to obtain the final co-added spectrum
(Fig.~\ref{fig_specco}).
The analysis presented is based on this final co-added spectrum.
Equivalent widths (Table~\ref{tab_wco}) 
were measured in various ways using the
tools available in IRAF/SPLOT, and derived column densities are
presented in Table~\ref{tab_logn}, and
a log of the observations is presented in Table~\ref{tab_log}.

\section{Molecular data}
\label{sec_molley}

The software program MOLLEY (Paper~III) has been modified and expanded to compute
the molecular parameters  (line position and line strengths)
for the  C$_{2}$ \phsys~ (Phillips) system  bands
and the CO \cosys~ rovibrational bands (and their isotopes).
Given the rotational temperature $T_{\rm rot}$, 
integrated molecular column density $\log N_{\rm int}$, Doppler
broadening parameter ($b$), and spectral resolution
$R=\lambda/\delta\lambda=\sigma/\delta\sigma$, a synthetic spectrum
can be computed.

\subsection{C$_{2}$}
\label{sec_molleyc2}

C$_{2}$
wavenumbers are computed from the molecular constants of
Marenin \& Johnson (\cite{mareninjohnson})
and agree very well  with the observed wavenumbers by 
Chauville \& Maillard (\cite{chauvillemaillard}).
Wavenumbers for
the isotopes were computed using 
the standard relations for the mass dependence of the various
molecular constants (c.f. Bernath~\cite{bernath})
and the molecular constants for C$_{2}$.
Wavelengths for C$^{13}$C 
(Table~\ref{tab_wcc}) have been computed
from the wavenumbers by Amiot \& Verges (\cite{amiotverges})
as measured by Fourier Spectroscopy.
Conversion from wavenumber (cm$^{-1}$ in vacuum) to wavelength (\AA~ in air)
was made using the index of refraction of standard
air as given by Morton (\cite{morton}).

C$_{2}$ band oscillator strengths for the Phillips
system have been determined experimentally, primarily from
measurements of the radiative lifetimes of vibrational levels
of the A$^{1}\Pi_{u}$ state, and estimated theoretically
from quantum chemistry calculations. Unfortunately, experiment
and theory are not in completely satisfactory agreement, see
for example, the review by Lambert et al. (\cite{lambertetal}).

Calculations appear to have converged upon a consistent set
of oscillator strengths and radiative lifetimes. Langhoff
et al. (\cite{langhoffetal}) estimate that their predicted
radiative lifetimes are accurate to about 5\%. A similar
accuracy surely applies to the band oscillator strengths. In
our study, we combine observations of lines from the four bands
1-0, 2-0, 3-0, and 4-0 for which the predictions are 
$f_{(1-0)}=0.00238$ and $f_{(2-0)}=0.00144$ by 
Langhoff et al. (\cite{langhoffetal}),
$f_{(3-0)}=0.0006672$ by Langhoff  (\cite{langhoffpc}) 
and $f_{(4-0)}=0.000271$ by van Dishoeck (\cite{vandishoeck})
(adjusted to the
band oscillator strength ratios of Langhoff et al.).
Radiative lifetimes from laser pyrolysis and laser-induced
fluorescence 
(Bauer et al. \cite{baueretal1},\cite{baueretal2})  
are appreciably longer than the 
predicted lifetimes implying experimental oscillator
strengths that are much smaller than the above estimates. Recent
measurements by a different technique give results closer to the
theoretical values:
Erman \& Iwamae (\cite{ermaniwamae}) measure, for example, 
a lifetime of $7.1\pm1.1~\mu$s for vibrational level
$v^{\prime}=4$ for which Langhoff et al.'s prediction
is $6.8~\mu$s but measurements of $11.1\pm1.1$ and $10.7\pm2.0~\mu$s
were reported by 
Bauer et al. (\cite{baueretal1})
and Bauer et al. (\cite{baueretal2}) respectively. In light of the 
consistent  theoretical results (see Langhoff et al. for references),
Erman \& Iwamae's new experiments, and the estimates derived
from observations of interstellar C$_{2}$ absorption
lines (Lambert et al. \cite{lambertetal}), 
we adopt Langhoff et al.'s predictions and their assessment
of their accuracy in our analysis.

The adopted method to compute line strengths from the
band oscillator strength is extensively
discussed in Paper~I and II, and we refer the interested reader
to that paper for all details.
In computing the line strength for the transitions
of the isotopes of C$_{2}$ (Table~\ref{tab_wcc} and MOLLEY)
we assume that the oscillator strength for
the C$^{13}$C band is the same as for the C$_{2}$ band.
Band heads of
$\nu_{(1-0)}=9854.0247$~cm$^{-1}$  
and $\nu_{(2-0)}=11413.8250$~cm$^{-1}$  have been taken
from Chauville \& Maillard (\cite{chauvillemaillard})
and for $\nu_{(3-0)}=12947.81$~cm$^{-1}$   
and $\nu_{(4-0)}=14459.01$~cm$^{-1}$ 
from  Ballik \& Ramsay (\cite{ballikramsay}).

C$_{2}$ is a homo-nuclear molecule, while
C$^{13}$C is a  hetero-nuclear molecule.
For C$_{2}$ only even $J^{\prime\prime}$ levels exist
($J^{\prime\prime}=0,2,4,6...$), 
while odd and even levels exist for  C$^{13}$C 
($J^{\prime\prime}=0,1,2,3...$).
The electric dipole moment for C$_{2}$ is strictly zero and therefore
pure rotational electric dipole transitions ($\Delta J = \pm 1$)
cannot occur. Electric
quadrupole transitions ($\Delta J = \pm 2$) may occur but with an extremely
low probability. The C$^{13}$C molecule has a very weak
electric dipole moment from failure of the Born-Oppenheimer
approximation and thus pure rotational transitions are
permitted, albeit with a low probability.
It is important to realize that the absence
of an infrared and sub-millimeter spectrum deprives C$_{2}$ 
of a cooling mechanism and the molecule is excited by the
stellar and interstellar radiation fields to supra-thermal temperature
($T_{\rm rot} \geq T_{\rm kin}$). C$^{13}$C
on the other hand has a small dipole moment,
and may cool.
The population distribution for the two molecules,
C$_{2}$ and C$^{13}$C, is therefore not necessarily the same
and the  rotational temperature
for C$^{13}$C may be lower than for C$_{2}$.
By studying these two molecules simultaneously, potentially
important information on the conditions
within the CSE could be extracted.

\subsection{CO}

CO wavenumbers are computed from the Dunham coefficients
given by Farrenq et al. (\cite{farrenqetal}).
Wavenumbers for Table~\ref{tab_wco} 
are after Pollock et al. (\cite{pollocketal}).

Oscillator strength for CO  have been computed
after Goorvitch \& Chackerian (\cite{goorvitcha},\cite{goorvitchb})
which have a band oscillator strength for CO of 
$f_{(2-0)}=0.8957\times 10^{-7}$. Oscillator strength 
of the 2-0 band determined
by Kirby-Docken \& Liu (\cite{kirbydockenliu78}) is  85\% 
of that of Goorvitch \& Chackerian.
This leaves us with an absolute uncertainty of at most 
15~\% on the adopted $f_{(2-0)}$-value. 	

\section{Analysis}

The analysis presented in this paper is based on the spectra listed
in Table~\ref{tab_log} supplemented with previous data
presented in Paper~I (C$_{2}$), II (C$_{2}$), and III (CN and $^{13}$CN),
and unpublished data for the C$_{2}$ 4-0 band.
Use of this combined set of data assumes that there has  been no
change of the excitation conditions of the molecules
in the interval of a few years.

The identification of the C$^{13}$C 1-0  and the 
CO 2-0 bands yield heliocentric velocities of
the line forming region of $78.5\pm0.4$ and 
$78.7\pm1.1$~km~s$^{-1}$. Within the errors, all velocities
of circumstellar molecular absorption lines
towards HD\,56126 
(C$_{2}$  : $76.6\pm0.2$;
C$^{13}$C : $78.5\pm0.4$; 
CN        : $77.5\pm0.5$;
$^{13}$CN : $76.7\pm1.8$;
CO        : $78.7\pm1.1$~km~s$^{-1}$) yield the same heliocentric velocity 
for the CSE of $v_{\rm CSE}=77.6\pm0.4$~km~s$^{-1}$
and successively an expansion velocity of 
$v_{\rm exp}=8.0\pm0.6$~km~s$^{-1}$.
The equality of $v_{\rm CSE}$ for the various species suggests that
their regions of residence overlap considerably.

\subsection{Doppler $b$-parameter}
\label{sec_bpar}

In order to derive the column density from the equivalent width
of a line, one has to take optical depth effects into account.
The C$^{13}$C and the CO lines are weak and optically
thin and their derived column densities
(Table~\ref{tab_logn}) are practically independent of
the adopted $b$-value. However, many of the C$_{2}$ lines are optical thick
and an accurate determination of the $b$-value is required to
computed accurate column densities.

Optical depth effects are quantified by the Doppler $b$-parameter
of the line absorption coefficient that determines the extension
of the curve of growth beyond the weak-line limit. To determine
the shape of the curve of growth and, hence, the $b$-parameter,
lines from a single vibrational band may be treated as a single dataset.
Such lines have accurate relative $f$-values set by the rotational line strengths
(H\"onl-London factors). Lines from a given lower 
($J^{\prime\prime}$) level may be combined immediately to form a part 
of the curve of growth. This curve for the C$_{2}$ Phillips bands
may consist of up to twelve  measured lines:
a P, Q, and a R line from each of three 1-0, 2-0, 3-0, and 4-0 
bands. This latter
step requires that the relative $f$-values of the three bands be known.
Sets of lines from different $J^{\prime\prime}$ levels may be combined after
determining the ratio of the column densities of the 
$J^{\prime\prime}$ levels or equivalently the rotational 
excitation temperature.

In Paper~III, circumstellar lines of the CN Red system were combined
to yield the $b$-parameter  of $b=0.51\pm0.04$~km~s$^{-1}$. 
The derived CN curve of growth is well definied. Particularly
worthy to note is the fact that the line sample for almost 
every rotational level spans almost the entire range of equivalent
width from weak unsaturated lines to saturated lines on the shoulder
of the curve of growth. Thanks to this circumstance, the derivation
of the $b$-parameter and the rotational populations 
are effectively independent and lead to accurate results. Furthermore,
the weak $^{13}$CN lines may be compared with rather similar weak
CN lines of the same rotational levels such that the
CN/$^{13}$CN ratio is independent of the derived $b$-parameter.
The ratio is dependent on the adopted ratio of $f$-values for the Violet
and Red system bands. (The derived $b$-parameter is dependent on
the relative $f$-values of the CN Red system bands.)

The C$_{2}$ Phillips system 1-0, 2-0, 3-0, and 4-0 bands do not
provide such a happy selection of lines. Lines from the four bands
from a given lower ($J^{\prime\prime}$) level do not span a large
range in equivalent widths. The line selection does not run from
weak to saturated lines. This necessarity means that the $b$-parameter
cannot be determined independently of the rotational population.
It should also be noted that, except for limited measurements
of the C$_{2}$ 1-0 lines, the equivalent widths comes from WHT
spectra. Comparision for CN lines show that the McDonald
spectra provide more accurate equivalent widths then those from 
the available WHT spectra.

\begin{figure*} 
\centerline{\hbox{\psfig{figure=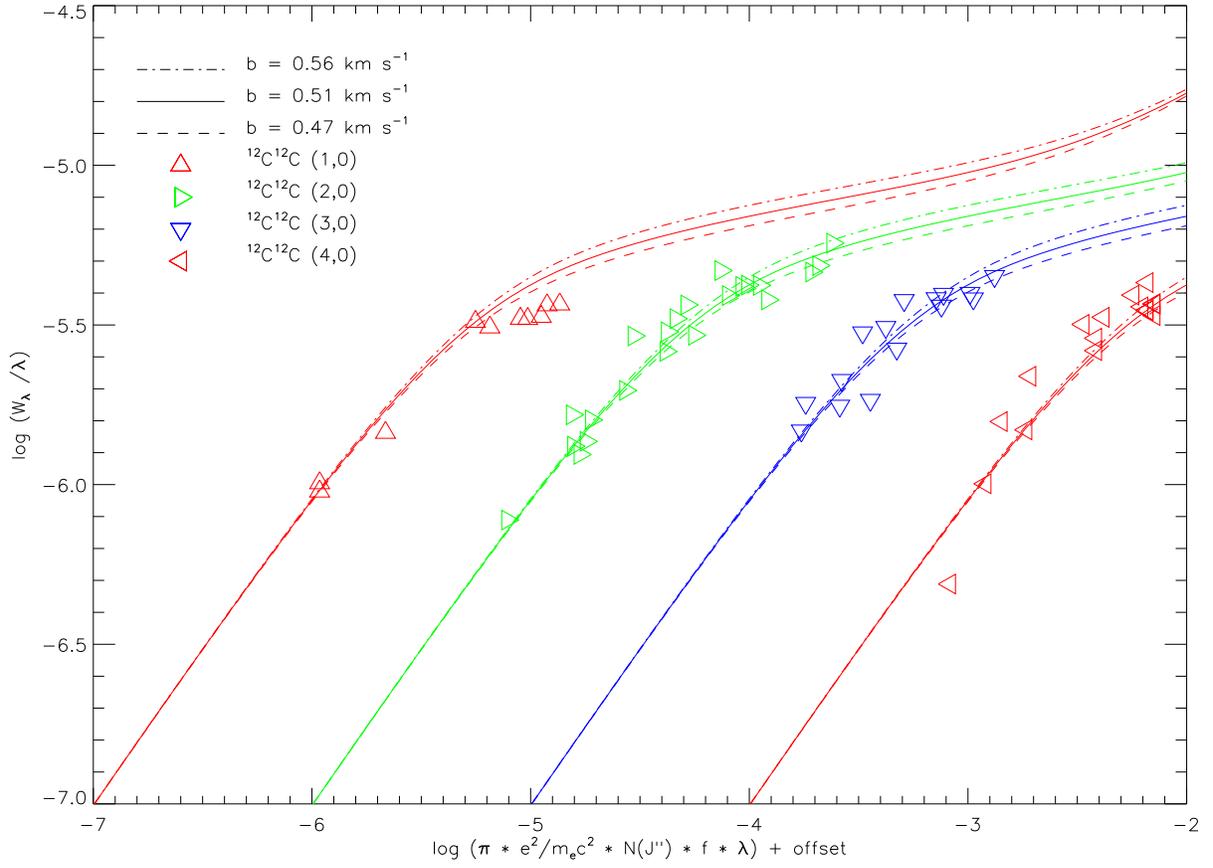,width=\textwidth,angle=90}}}
\caption{Curve of growth for $b=0.51$~km~s$^{-1}$ 
and the column densities of Table~\ref{tab_logn}.
The C$_{2}$ data points are labeled with an open triangle
(1-0 up; 2-0 right; 3-0 down; 4-0 left). 
The C$_{2}$ bands are offset by multiples of 1 dex 
in the $x$-direction.
The $f$-values have been multiplied by 0.62 1-0, 0.98 2-0,
1.53 3-0, and 2.26 4-0 
to obtain an improved fit (for details see Sec.~\ref{sec_bpar}).}
\label{fig_cofg}
\end{figure*}

In light of these limitations, we elected to assume that the 
appropriate $b$-parameter for C$_{2}$ was the well-determined
value  found for CN. Curves of growth for the individual C$_{2}$ 
bands are shows in Fig.~\ref{fig_cofg} where the column densities
are defined using all available lines. This shows clearly that
the observed curve of growths of the bands needs to be shifted
systematically relative to the theoretical curves. As noted above,
this inconsistency between observed and theoretical curves of growth
was not found for the CN Red system bands.

Several potential explanations for this problem affecting
the C$_{2}$ curves of growth have been considered.
Could the relative $f$-values be at fault? If the 
$f$-values are considered as the adjustable parameter, the 
necessary adjustments
corresponds to multiplicative factors of 0.62, 0.98, 1.53, and 2.26 for
the 1-0, 2-0, 3-0, and 4-0 bands respectively. Recall that the adopted 
theoretical $f$-values have been estimated to be accurate to about 
5~\%. It is instructive to compare relative $f$-values: the requirement
that lines from the four bands be forced to fit a single curve of 
growth requires ratios of $f_{(2-0)}/f_{(1-0)}=1.1$ and 
$f_{(3-0)}/f_{(1-0)}=0.8$. In contrast, the 
theoretical estimates of the same ratios are 0.61 and 0.28 from
Langhoff et al. with very similar ratios from other theoretical 
calculations; for example, van Dishoeck's (\cite{vandishoeck})
$f$-values correspond to the ratios 0.59 and 0.27. Earlier 
calculations by Theodorakopoulos et al. (\cite{theodorakopoulos})
calculations give the ratios of 0.64 and 0.31. In general, 
absolute values of the $f$-values are more sensitive to the details
of a calculation than are ratios of the $f$-values that depend
primarily on the well determined potential energy curves
and secondarily on the variations of the electronic 
transition moment with inter-nuclear separation. In other words,
it is not surprising that Theodorakopoulos et al.'s
$f$-values for the 2-0 band is 29~\% larger than Langhoff et al.`s
but that the above sets of $f$-values ratios from the same pair of
calculations are identical to within 5-10~\%.
It is worth noting too that observations of interstellar C$_{2}$
lines toward $\zeta$ Oph. give the ratio
$f_{(2-0)}/f_{(3-0)}=2.1\pm0.4$ (van Dishoeck \& Black \cite{vandishoeckblack})
in good agreement with the predictions from Langhoff et al. of 2.15, and
in sharp disagreement with the ratio of 1.4 from fitting HD\,56126's
lines to a single curve of growth.
Our assessment of the theoretical $f$-values for the Phillips system
is that their uncertainties do {\it not} allow curves of growth
for the individual bands  to be shifted as here required.

Could emission affect the equivalent widths?
If unresolved emission from the circumstellar shell fills in 
the absorption provided by C$_{2}$ molecules along the line of
sight to the star, the theoretical curves of growth for
a uniform slab will be inappropriate.
If the emission contribution declines in the
order 1-0, 2-0, 3-0, and 4-0, the apparent inconsistencies in the 
column densities could be explained. It is certainly the case that, 
if the circumstellar shell is spherically symmetric and large with
respect to the star, one expects a P-Cygni profile with blue-shifted
absorption and emission to the red of the absorption.
(Reddy et al. (\cite{reddybakkerhrivnak}) observed
that the circumstellar C$_{2}$ Phillips lines of
the post-AGB star IRAS\,07431+1115 exhibit P-Cygni profiles
while
Cohen \& Kuhi (\cite{cohenkuhi}) found that the absorption component of C$_{2}$
towards the Egg Nebulae was present in the polarized light
(reflected by the nebulae), while the emission component
was unpolarized.)
Since the expansion velocity is 8.0~km~s$^{-1}$, the emission
and absorption would be fully resolved at our resolution such that dilution
of the absorption should be very small. Such emission is not seen
in our spectra. We conclude that the shell does not meet the conditions
of large and/or spherically symmetric.
A special geometry must be 
provided such that emission from regions off the line of sight to the star
provide sufficient flux to fill in the lines with the additional
proviso that emission not dominate entirely.

Transition probabilities of Phillips system bands do suggest that
emission might account for the apparent inconsistencies in the column
densities. Branching ratios, $p_{v^{\prime} - v^{\prime\prime}}$,
for emission from the levels $v^{\prime}$ of the upper state
may be calculated from transition probabilities, 
A$_{v^{\prime} - v^{\prime\prime}}$, given by van Dishoeck
(\cite{vandishoeck}). We find that $p_{(1-0)}=82$~\%, that is 82~\%
of the radiative decays from $v^{\prime}=1$ are in the 1-0 band. For the
other observed bands, $p_{(2-0)}=55$~\%,
$p_{(3-0)}=28$~\%, and $p_{(4-0)}=12$~\%. Emission
rates of these bands depend also on the pumping rates
from $v^{\prime\prime}=0$ in the  X$^{1}\Sigma^{+}_{g}$ state.
These rates should be fairly similar to $v^{\prime}=$ 1, 2, 3, and 4.
Then, it is apparent that contamination of the absorption lines by emission
should be most serious for the 1-0 band and least for the 4-0 band.
This is in the correct sense to account for the apparent inconsistencies
in the column densities.
It follows that, if this is the correct explanation that C$_{2}$ bands
should also be seen in emission. for example, the 3-1 band is
expected to have twice the emission strength of the 3-0 band.
We looked for emission of the 3-1 band in the  available
spectra which cover this region, no such emission (nor absorption)
was found to be present.
A simpler and stronger argument against the ``emission'' hypothesis
is that no such effects are seen in the CN Red system lines where
the run of $p_{v^{\prime} - v^{\prime\prime}}$ from 73~\% to 7~\%
for 1-0 to 4-0 is similar to the run for the Phillips bands.

Could the CN $b$-parameter be inappropriate for C$_{2}$?
The CN $b$-parameter is dominated by contributions from small
scale turbulence or a velocity gradient. Since CN and C$_{2}$ seem
likely to reside in similar layers of the shell, it seemed
reasonable to suppose that the molecules have the same
$b$-parameter. It is, however, the case that a lower value,
$b=0.31\pm0.05$~km~s$^{-1}$ put all lines on a single curve
of growth whith adoption of the theoretical $f$-values. We
cannot reject this lower $b$-parameter directly but note (see below)
that it gives an unusually high isotopic ratio of C/$^{13}$C$\geq 300$.

Could the equivalent widths be in error ?
A noticeable difference between the CN and C$_{2}$ curves of growth
is the larger scatter of the latter about the fitted theoretical
curve of growth. We suspect that this reflects the fact that the 
CN equivalent widths are measured of higher resolution and
higher signal-to-nose ratio spectra.  Then the vast majority of
the C$_{2}$ lines, in particular, the C$_{2}$ 1-0 lines are in a 
ratter noisy region of the available spectra. Quite possibly, the 
C$_{2}$ lines are subject to systematic errors. One contributing
factor may be that the wings of the strongest lines make a significant
contribution of the total equivalent width.
For circumstellar lines (with a low $b$-value), the intrinsic
spectrum (as observed at infinite spectral resolution) reaches zero intensity
in the core of the profile. Line wings are very extended (several times the
$FWHM$ of the line) and represent a significant fraction of the column density.
These shallow wings are unfortunately very hard to measure and in most cases
we did a Gaussian fit on the profile to obtain the equivalent width and assumed
that the contribution of the wing is irrelevant.
Tests on the difference between a Gaussian and Voigt profile 
shows indeed that this could account for a part of the inconsistensies.
The stronger the line, the more our measured equivalent widths are in error.
Unfortunately there
is no easy way to measure the line wings unless we obtain additional very high
signal-to-noise ratio spectra at very high-resolution ($R\geq 350\,000$).

In conclusion, the inconsistencies in the derived column densities 
are most likely a result of the fact that we did not
take the extended wings of the absorption profiles into account. Hence, for the 
saturated lines, our measured equivalent widths are too low.

Fig.~\ref{fig_cofg} shows the curve of growth of C$_{2}$ Phillips lines
towards HD\,56126. The $f$-values for the Phillips bands have
been multiplied by 0.62 1-0, 0.98 2-0, 1.53 3-0, and 2.26 4-0 to
improve the fit, and
the over-plotted curve of growth is for our preferred value
of $b=0.51\pm0.05$~km~s$^{-1}$. Two other curve of growth have been over-plotted
which have a $b$-value marking the error bar of our $b$ determination.
By changing the relative
$f$-values to obtain consistent column densities, we find that
this introduces a shift of $b$ of 0.05~km~s$^{-1}$ towards
lower $b$-values. We therefore
conclude that our analysis is not severely affected, but that
an increased accuracy can only be reached by taking the 
(presumed) emission component into account.

In deriving C$_{2}$ column densities, we suppose that the theoretical
$f$-values and $b=0.51$~km~s$^{-1}$ are correct, and attribute the offset
from the theoretical curve of growth to errors affecting the measured
equivalent widths. For the latter reason, we give low weight
to the strong C$_{2}$ 1-0 lines.

Column densities (Table~\ref{tab_logn})
for C$_{2}$, C$^{13}$C, CN, $^{13}$CN, and CO have
been computed by means of a curve of growth for $b=0.51$~km~s$^{-1}$ 
and using the $f$-values given in Sec.~\ref{sec_molley}.

\subsection{Rotational diagram and column densities}

A rotational diagram has been constructed,
using the $f$-values given in Sec.~\ref{sec_molley}, for
C$_{2}$, C$^{13}$C, and CO (Fig.~\ref{fig_rot}) using the column 
densities presented in Table~\ref{tab_logn}. For a molecule in
local thermodynamic equilibrium (LTE) for which the population
distribution of its energy levels can be described by
a Boltzmann distribution,  the rotational diagram is
a straight line. The rotational temperature is given by
the negative of the inverse of the slope of the line.
From previous work (see for details and references Paper I, II, III)
we know that C$_{2}$ and CN are not in LTE,
their rotational diagram are not linear.
The curvature for C$_{2}$ in Fig.~\ref{fig_rot} suggests that
this molecule is not in LTE. For the other molecules, too few $J^{\prime\prime}$
levels are probed to make a similar claim.

To quantify the  excitation of the molecule we make a linear fit to the rotational
diagram to obtain an effective rotational temperature. We find
$T_{\rm rot} = 328\pm37$~K, $\log N_{\rm int}=15.34\pm0.10$~cm$^{-2}$,
$T_{\rm rot} = 256\pm30$~K, $\log N_{\rm int}=13.79\pm0.12$~cm$^{-2}$, and
$T_{\rm rot} =  51\pm37$~K, $\log N_{\rm int}=18.12\pm0.13$~cm$^{-2}$ 
for C$_{2}$, C$^{13}$C, and CO respectively (see also Table~\ref{tab_logn}).
The column densities from the observed levels are not always a good measure
for the total column densities. This is best demonstrated by looking
at C$^{13}$C. With a rotational temperature of 256~K, the level that
is most populated is $J^{\prime\prime}=7$. Since we observed only
the Q-Branch with $J^{\prime\prime}$ lower than 9, we miss about
50\% of the column density. By assuming a constant rotational temperature
and extrapolating to the unobserved levels, we can compute the 
total C$^{13}$C column density. Another problem is encountered
for C$_{2}$. Since this molecule is not in LTE, the integrated column
density will be different from the total observed column density.

\section{Discussion}

\begin{figure*} 
\centerline{\hbox{\psfig{figure=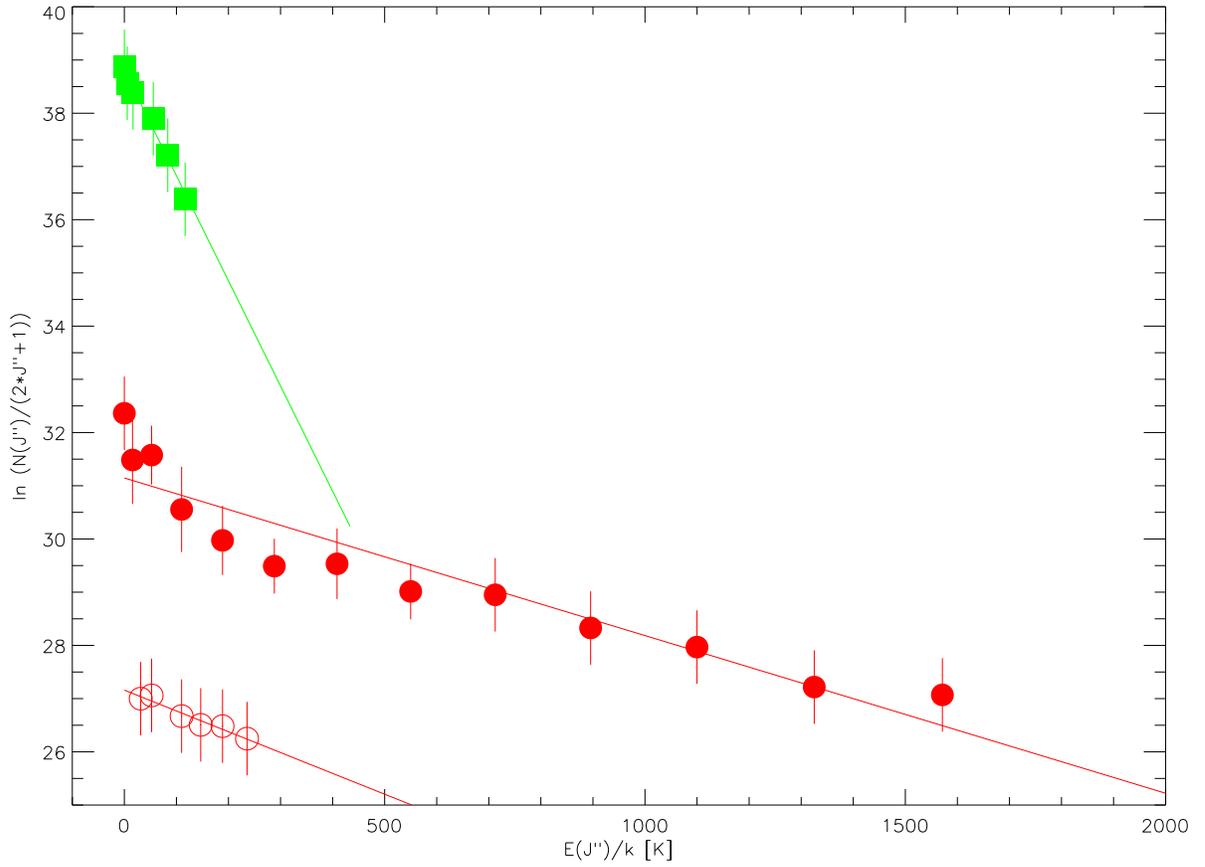,width=\textwidth,angle=90}}}
\caption{Rotational diagram for C$_{2}$
(filled circles) and C$^{13}$C (open circles) Phillips 
system bands, and the CO (filled squares) first-overtone band towards HD\,56126. 
Column densities  and errors used are listed in Table~\ref{tab_logn}.
Note that the curvature for the C$_{2}$ data points is real
and is attributed to non-LTE effects.}
\label{fig_rot}
\end{figure*}

The rotational temperatures of C$^{13}$C and CO are well constrained
by a linear fit to their rotational curves (Fig.~\ref{fig_rot}).  
Within the error C$_{2}$ can be fitted to the same rotational temperature
as C$^{13}$C, but 
it is clear that the C$_{2}$ rotational curve  has a depression from
a linear curve near $E(J^{\prime\prime})/k=300$~K. 
For lower energies (lower J$^{\prime\prime}$), 
the curve is steeper
which translates in a lower rotational temperature, while for higher energies
(higher J$^{\prime\prime}$),
the curve is flatter which translates in a higher rotational temperature.
This effect
has been noted before (Paper~I, II) and is characteristic for a molecule
which is not in LTE and has a rotational temperature in excess of 
the kinetic temperature of the gas. C$_{2}$ is a homo-nuclear molecule
and has therefore no permanent dipole moment. Selection rules do not allow
rovibrational (infrared) and rotational (sub-millimeter) transitions.
The molecule absorbs energy from the infrared to the ultraviolet
(the observed molecular bands), but has no strong cooling mechanism.
The result is that the molecule is rotationally excited 
above the kinetic temperature.
An overall fit to the curve yields $T_{\rm rot}=328\pm37$~K. 
C$^{13}$C has a very small electric dipole moment
and very weak allowed rovibrational and rotational transitions allow the
molecule to cool. The possibly lower rotational temperature 
for C$^{13}$C of $T_{\rm rot}=256\pm30$~K is consistent with this model.
CO has a  permanent dipole moment 
($0.1222$~Debye, for references see Kirby-Docken \& Liu \cite{kirbydockenliu77}).
It can effectively
cool by emitting a photon in the infrared or sub-millimeter. It is thus not 
unexpected that we find a rotational temperature lower than that for C$_{2}$
and C$^{13}$C. CN, with a permanent dipole moment of $1.45\pm0.08$~Debye
(Thomson \& Dalby \cite{thomsondalby}),
cools more efficiently and has a lower rotational temperature
of $T_{\rm rot}=11.5\pm0.6$~K.
Besides the difference in dipole moment, there are second order 
factors which should be considered in order to understand the 
excitation of the molecule. Since each molecules has its own characteristic
absorption spectrum, different molecules absorb different parts of the stellar spectrum.
The energy distribution of the central stars,
the star's photospheric spectrum, and the reddening between the star
and the line forming region should therefore be considered. 
Clearly, it would be of great interest
to model the excitation of the five molecules detected and extract
information about the conditions within the CSE from their excitation.

Because of the difference in rotational temperature between
C$_{2}$ and C$^{13}$C, and the fact that C$^{13}$C
has twice as many $J^{\prime\prime}$ levels available, the C$_{2}$/C$^{13}$C
can not be determined for each  $J^{\prime\prime}$ 
level independently. Instead
the only acceptable way is by integrating the population over all
available levels. The unobserved levels for C$^{13}$C contain a significant
fraction of the total column density and the total C$^{13}$C 
column density can therefore
only be obtained by computing a rotational temperature and extrapolating to
the unobserved energy levels (labeled as $\log N_{\rm int}$ in 
Table~\ref{tab_logn}).
Using the integrated column densities for C$_{2}$ and C$^{13}$C we find
C$_{2}$/C$^{13}$C$=36\pm13$. 

The conversion from C$_{2}$/C$^{13}$C to C/$^{13}$C is rather
straightforward. Under chemical equilibrium conditions,
C$^{13}$C is formed twice as easily as C$_{2}$ (Tatum \cite{tatum}).
This can be best demonstrated with Tatum's example. In 
a draw there are equal number of white and black socks 
($^{12}$C and $^{13}$C atoms to make
C$_{2}$H$_{2}$ $\rightarrow$ C$_{2}$H $\rightarrow$ C$_{2}$). 
By taking at random two socks out
of the draw, the possible combinations are 
(white,white), (white,black), (black,white), and (black,black). 
Since (white,black) is the same as
(black,white) the probability is 1:2:1 for a white, mixed, and black pair
of socks. Hence the equilibrium abundance of the hetero-nuclear molecule must
be twice that of the homo-nuclear molecule. Excluding any
other reaction this leads to C/$^{13}$C$=2 \times$C$_{2}$/C$^{13}$C=$72\pm26$.

CN is a polar molecule with a permanent dipole moment.
The rate coefficient for the isotopic exchange reaction of CN with
C$^{+}$ is as high as $10^{-7}$~cm$^{-3}$~s$^{-1}$
(Adams et al. \cite{adamsetal}). Since $^{13}$CN has a 31~K lower
zero-point energy than CN, the reaction making $^{13}$CN from
CN is exothermic. In paper~III we constructed a simple chemical
model of the CSE of HD\,56126 and found 
that for C$_{2}$/$^{13}$CN$=38\pm2$ we get C/$^{13}$C$\simeq 67$.
C$_{2}$ is a symmetric molecule possessing no permanent
dipole moment, theory predicts a rate
coefficient which is typically 
$10^{-9}$~cm$^{-3}$~s$^{-1}$. Since C$^{13}$C has a 25~K lower
zero-point energy than C$_{2}$ 
the reaction making C$^{13}$C from C$_{2}$ is exothermic.
Since the zero-point energy difference for
C$^{13}$C and $^{13}$CN to C$_{2}$ and CN respectively are about the same,
and C$_{2}$ and CN have about the same abundance with distance
from the star,
we can use the chemical model of Paper~III (their Fig.~6.d.)
to assess the importance of the isotopic exchange reaction for C$_{2}$.
We find that our model predicts C$_{2}$/C$^{13}$C=60 for
C/$^{13}$C=67. Taking into account the many uncertainties of this model
(C$^{+}$ abundance, kinetic temperature etc.) 
we argue that within the errors of our computations,
the isotopic exchange reaction does not significantly alter the 
C$_{2}$/C$^{13}$C ratio in the circumstellar shell surrounding HD\,56126.

Given these argument for C$_{2}$ and CN, we argue that
the intrinsic carbon isotope ratio is well constrained
by the C$_{2}$/C$^{13}$C ratio as C/$^{13}$C=$72\pm16$,
and that a rate coefficient 
for the CN isotopic exchange reaction of $10^{-7}$~cm$^{-3}$~s$^{-1}$ 
is consistent with the observed CN/$^{13}$CN ratio.

The prototype of the AGB stars, IRC~+10216/CW Leo,
is a massive, highly-evolved carbon star 
with $3 M_{\odot} \leq M_{\rm ZAMS} \leq 5 M_{\odot}$
(Gu\'{e}lin et al. \cite{guelinetal}). For IRC~+10216 the isotope
ratios are well determined and give an estimate of the ratios
which one might expect to detect for a carbon-rich post-AGB star
like HD~56126: C/$^{13}$C=44$^{+3}_{-3}$.
(see Forestini \& Charbonnel  \cite{forestinicharbonnel} for  an overview). 
Our estimate of the C/$^{13}$C ratio for HD~56126's shell is 
consistent with that of IRC~+10216. Results for circumstellar
shells around four other carbon stars were provided by 
Kahane et al. (\cite{kahane}): C/$^{13}C \leq 60$.
These too suggest that HD~56126 is not exceptional.

Combining our results from Paper~III and this paper yields
a column density ratio  of
CO/(C$_{2}$+C$^{13}$C)$=606\pm230$ and
CO/(CN+$^{13}$CN)$=475\pm175$.
These column density ratios do not necessarily reflect 
abundance density ratios. CO is abundant throughout the whole
CSE, while C$_{2}$ and CN are only present in a shell
within the CSE where interstellar photons photo-dissociate
HCN to CN and  C$_{2}$H$_{2}$ via C$_{2}$H to C$_{2}$ 
(Cherchneff et al. \cite{cherchneff}).
Column density ratios of CO to C$_{2}$ and CN (excluding their isotopes)
have been measured in several ways. Olofsson (\cite{olofsson}) lists
abundances for a large range of molecules detected in the shell
surrounding AGB stars (although most molecules are only detected
in IRC\,+10216). From his work we find CO/C$_{2}=1 \times 10^{3}$
and CO/CN$=1 \times 10^{3}$. Bachiller et al. (\cite{bachilleretal})
present abundance ratios for, among others,  CRL~2688 (Egg Nebulae) 
obtained  from sub-millimeter line emission observations 
CO/CN=909. Within a factor of two these numbers are consistent with
what we find for HD\,56126.
Bachiller et al. suggest that the CO/$^{13}$CO$\simeq 20$ likely reflects
the C/$^{13}$C for the planetary nebulae. If so, their C/$^{13}$C
is a factor three lower than ours. This could suggest that the objects
observed by Bachiller et al. are not the precursors of HD\,56126 but
possible PN's with higher mass progenitor ($M_{\ast} \geq 2$~M$_{\odot}$)
while the progenitor of HD\,56126  has ($M_{\ast} \leq 2$~M$_{\odot}$).

A model for the C$_{2}$ molecule has been constructed by van Dishoeck \& Black
(\cite{vandishoeckblackmodel}) which takes into account collisional and
radiative excitation and de-excitation. For $T_{\rm kin}=25$~K we find
$S=n_{\rm c} \sigma / I=3.3 \pm 1.0 \times 10^{-14}$~cm$^{-1}$
(see also Bakker et al. \cite{bakkeredinburg}).
It is curious that these estimates of $T_{\rm kin}$ and $S$
are quite similar to those derived from C$_{2}$ molecules in
diffuse interstellar clouds. This is surprising as in Paper~II we
estimated that the circumstellar C$_{2}$ molecules are pumped
largely by the stellar radiation field.
From a fit to the rotational population, 
$\sigma=7.8 \times 10^{-16}$~cm$^{2}$ as the collisional cross-section
for C$_{2}$-H$_{2}$ (Phillips \cite{phillips}), and 
$I(r=10^{16}~{\rm cm}) \simeq 4.1 \times 10^{5}$ the ratio of
the stellar  radiation field  relative to the
standard interstellar radiation field, we find
$n_{\rm c}=n ({\rm H}) + n({\rm H_{2}})=1.7 \times 10^{7}$ 
as the number of collisional partners for C$_{2}$. At 
$r=10^{16}$~cm this translates in
$\dot M \simeq 2.5 \times 10^{-4}$~M$_{\odot}$~yr$^{-1}$. This 
mass-loss rate as derived from modeling the C$_{2}$ excitation
gives a result comparable to the mass-loss rate derived
using other techniques.

Based on this work we can make some suggestions to continue this study.
A critical detection would be that of the quadrupole transitions
of H$_{2}$ at 2.2$\mu$m and $^{13}$CO rovibrational bands at 4.6~$\mu$m. 
This would allow to determine the column density
ratio of the detected molecules relative to the most abundant specie (H$_{2}$)
and the CO/$^{13}$CO ratio. 
Secondly, more C2CN stars should be studied to determine the C$_{2}$/C$^{13}$C
and CN/C$^{13}$N ratio to get good statistics.
The most ambitious project would be to model the excitation of
all molecules simultaneously in order to obtain the physical conditions
of the CSE (densities, radiation field, extinction etc.).

\acknowledgments
The authors acknowledge the support of
the National Science Foundation (Grant No.
AST-9618414) and the Robert A. Welch Foundation of Houston, Texas.
This research has made use of the Simbad database, operated at
CDS, Strasbourg, France, the ADS service, and IRAF. We thank
the IRTF staff and John Rayner, for their support in obtaining
CSHELL spectra.

\clearpage

\begin{table*} 
\caption{Log of observations of C$^{13}$C 1-0 and CO 2-0.}
\label{tab_log}
\begin{tabular}{llllrl}
\hline
\hline
Date         &$HJD^{a}$ &Exp. time [s] &$\lambda^{b}$ [\AA]
                                                 &$SNR$     &Remark\\
\hline
             &          &              &         &    &\\
\multicolumn{3}{l}{HD\,56126}                     &    &\\
10 Jan.  1998&2450823.68&$12\times1800$          &8876& 20&C$^{13}$C 1-0 Q-branch\\
11 Jan.  1998&2450824.68&$12\times1800$          &8876& 20&C$^{13}$C 1-0 Q-branch\\
12 Jan.  1998&2450825.65&$12\times1800$          &8876& 20&C$^{13}$C 1-0 Q-branch\\
\multicolumn{3}{l}{$\beta$ Ori}                  &    &   &\\
10 Jan.  1998&2450823.64&$5\times  200$          &8876&273&...\\
11 Jan.  1998&2450824.65&$3\times  200$          &8876&230&...\\
\multicolumn{3}{l}{co-added HD\,56126/$\beta$ Ori}&8876& 85&C$^{13}$C 1-0 Q-branch\\
             &          &                        &    &   &\\
\hline
Date         &$HJD^{a}$ &Exp. time [s]       &$\sigma$ [cm$^{-1}$]
                                                  &$SNR$&Remark\\
\hline
             &          &                    &    &   &\\
\multicolumn{3}{l}{HD\,56126}                 &    &   &\\
22 Feb. 1998&2450866.718&$34\times 300$      &4281& 87&CO 2-0 R($J^{\prime\prime}$)= 4,5,6 \\
23 Feb. 1998&2450867.704&$32\times 300$      &4266&100&CO 2-0 R($J^{\prime\prime}$)= 0,1,2 \\
\multicolumn{3}{l}{HR\,2763}                  &    &   &\\
22 Feb. 1998&2450866.770&$24\times  60$      &4281&163&CO 2-0 R($J^{\prime\prime}$)= 4,5,6 \\
23 Feb. 1998&2450867.756&$20\times  60$      &4266&711&CO 2-0 R($J^{\prime\prime}$)= 0,1,2 \\
\multicolumn{3}{l}{co-added HD\,56126/HR\,2763}&4281&170&CO 2-0 R($J^{\prime\prime}$)= 4,5,6 \\
\multicolumn{3}{l}{co-added HD\,56126/HR\,2763}&4266&180&CO 2-0 R($J^{\prime\prime}$)= 0,1,2 \\
            &           &                    &    &   &\\
\hline 
\hline
\end{tabular} \\
$^{a}$ heliocentric Julian date of first observation. \\
$^{b}$ central wavelength or wavenumber of relevant order. \\
\end{table*}

\clearpage

\begin{table*} 
\caption{C$^{13}$C Phillips system 1-0 band for 
the transitions with $J^{\prime\prime}\leq 9$ (lines are listed in order
of increasing wavelength). 
Derived column densities are given in Table~\ref{tab_logn}.}
\label{tab_wcc}
\begin{tabular}{llllll}
\hline
\hline
$B$($J^{\prime\prime}$) &$\lambda_{\rm rest}^{a,b}$ [\AA]&$\lambda_{\rm helio}$ [\AA]
                                 &$f_{J^{\prime},J^{\prime\prime}}^{c}\times10^{3}$  
                                                &$W_{\lambda}$ [m\AA]
                                                         & Remarks \\
      &                          &               &$\pm0.1$&$\pm0.3$ \\
\hline
      &           &           &     &   &         \\
R(7)  &10162.961l &10165.50   &0.713&0.6&tentative\\
R(6)  &10163.064l &10165.60   &0.731&1.6&tentative\\
R(8)  &10163.297c &10165.70   &0.699&2.1&tentative\\
R(5)  &10163.605l &10166.27   &0.756&2.2&tentative\\
R(9)  &10164.070l &10166.80   &0.688&2.8&tentative\\
R(4)  &10164.583c &...        &0.792&...&blended\\
R(3)  &10165.998c &10168.66   &0.848&2.7&...     \\
R(2)  &10167.851l &...        &0.950&...&not detected\\
R(1)  &10170.142c &...        &1.187&...&not detected\\
R(0)  &10172.779c &...        &2.380&...&not detected\\
Q(1)  &10176.478c &...        &1.186&...&not detected\\
Q(2)  &10177.352c &...        &1.186&...&blended \\
Q(3)  &10178.665c &10181.33   &1.186&4.0&...\\
Q(4)  &10180.412l &10183.08   &1.186&5.2&...\\
Q(5)  &10182.598l &10185.24   &1.186&2.7&... \\
P(2)  &10183.699c &...        &0.237&...&not detected\\
Q(6)  &10185.226l &10187.90   &1.185&5.1&...\\
P(3)  &10188.189c &...        &0.339&...&not detected\\
Q(7)  &10188.294l &10190.96   &1.185&5.0&...\\
Q(8)  &10191.803c &10194.48   &1.185&5.5&...\\
P(4)  &10193.123c &...        &0.395&...&not detected\\
Q(9)  &10195.751l &10198.39   &1.184&4.9&... \\
P(5)  &10198.450c &10201.15   &0.431&2.7&tentative\\
P(6)  &10204.313l &...        &0.455&...&not observed \\
P(7)  &10210.583l &...        &0.473&...&not observed \\
P(8)  &10217.309l &...        &0.487&...&not observed \\
P(9)  &10224.469l &...        &0.497&...&not observed \\
      &           &           &     &   &         \\
\hline
\hline
\end{tabular}\\
$^{a}$ wavelengths have been computed from the wavenumbers by 
Amiot \& Verges (\cite{amiotverges})
and the index of refraction of $n_{\rm air}=1.0002741$ 
(Morton \cite{morton}).\\
$^{b}$ appended to the wavelength; l: laboratory; c: computed.\\
$^{c}$ computed with $f_{(1-0)}=0.00238$. \\
$^{d}$ heliocentric velocity of the identified lines 
$v_{\rm helio}=78.5\pm0.4$~km~s$^{-1}$
(excluding tentative and blended lines).\\
\end{table*}

\clearpage

\begin{table*} 
\caption{CO 2-0 band for 
transitions with $J^{\prime\prime}\leq 9$.
Derived column densities are given in  Table~\ref{tab_logn}.}
\label{tab_wco}
\begin{tabular}{llllll}
\hline
\hline
$B$(J$^{\prime\prime}$) &$\sigma_{\rm rest}^{a,b}$ [cm$^{-1}$] 
       &$f_{J^{\prime},J^{\prime\prime}}^{c}\times10^{8}$  
       &$\sigma_{\rm obs}^{d}   $ [cm$^{-1}$] 	
       &$W_{\sigma}\times10^{3} $ [cm$^{-1}$]
       &Remarks \\
       &         
       &     
       &$\pm0.10$
       &$\pm0.10$
       &\\
\hline
          &         &     &       &    &        \\
R(0)      &4263.837l&8.957&4262.45&5.2 &...     \\
R(1)      &4267.542l&6.007&4266.14&7.1 &...     \\
R(2)      &4271.177l&5.439&4269.75&8.5 &...     \\
R(3)      &4274.741l&5.212&...    &... &not observed  \\
R(4)      &4278.234l&5.098&4276.83&8.7 &...     \\
R(5)      &4281.657l&5.036&4280.23&5.9 &blended    \\
R(6)      &4285.009l&5.002&4283.55&3.3 &blended    \\
R(7)      &4288.290l&4.985&...    &... &not observed  \\
R(8)      &4291.499l&4.979&...    &... &not observed  \\
R(9)      &4294.638l&4.981&...    &... &not observed  \\
          &         &     &       &    &        \\
\hline
\hline
\end{tabular}\\
$^{a}$ after Pollock et al. (\cite{pollocketal}). \\
$^{b}$ appended to the wavenumber; l: laboratory.\\
$^{c}$ computed with $f_{(2-0)}=8.95\times10^{-8}$. \\
$^{d}$ heliocentric velocity of the identified lines 
$v_{\rm helio}=78.7\pm1.1$~km~s$^{-1}$
(excluding blended lines).\\
\end{table*}

\clearpage

\begin{table*} 
\caption{Derived column densities and column density ratios based on a 
curve of growth analysis for a Doppler broadening 
parameter of $b=0.51$~km~s$^{-1}$ and the $f$-values given
in Sec.~\ref{sec_molley}.}
\label{tab_logn}
\begin{tabular}{lrlrlrlrlrl}
\hline
\hline                       &
\multicolumn{2}{c}{C$_{2}$  }&
\multicolumn{2}{c}{C$^{13}$C}&
\multicolumn{2}{c}{CN       }&
\multicolumn{2}{c}{$^{13}$CN}&
\multicolumn{2}{c}{CO       }\\
$J^{\prime\prime}$ or $N^{\prime\prime}$        &
no.$^{a}$&$ \log N(J^{\prime\prime})^{b}$       &
no.$^{a}$&$ \log N(J^{\prime\prime})^{b}$       &
no.$^{a}$&$ \log N(N^{\prime\prime})^{c}$       &
no.$^{a}$&$ \log N(N^{\prime\prime})^{c}$       & 
no.$^{a}$&$ \log N(J^{\prime\prime})^{b}$\\
                             &   
&[cm$^{-2}$]                 &
&[cm$^{-2}$]                 &
&[cm$^{-2}$]                 &
&[cm$^{-2}$]                 &
&[cm$^{-2}$]\\
\hline
             &   &              & &              &  &              &                & &                        \\
 0           &3  &$14.05\pm0.30$&0&...           &10&$14.80\pm0.17$&1&$13.22\pm0.10$&1&$16.89\pm0.30$          \\
 1           &...&...           &0&...           &12&$15.05\pm0.17$&2&$13.59\pm0.10$&1&$17.22\pm0.30$          \\
 2           &12 &$14.37\pm0.36$&0&...           &17&$14.85\pm0.16$&2&$13.16\pm0.10$&1&$17.37\pm0.30$          \\
 3           &...&...           &2&$12.57\pm0.30$&11&$14.39\pm0.13$&2&$12.44\pm0.10$&0&...                     \\
 4           &6  &$14.67\pm0.24$&1&$12.71\pm0.30$& 1&$13.78\pm0.10$&0&...           &1&$17.41\pm0.30$          \\
 5           &...&...           &0&...           & 1&$12.70\pm0.10$&0&...           &1&$17.20\pm0.30$          \\
 6           &7  &$14.38\pm0.35$&1&$12.70\pm0.30$& 0&...           &0&...           &1&$16.91\pm0.30$          \\
 7           &...&...           &1&$12.69\pm0.30$& 0&...           &0&...           &0&...                     \\
 8           &7  &$14.25\pm0.29$&1&$12.73\pm0.30$& 0&...           &0&...           &0&...                     \\
 9           &...&...           &1&$12.68\pm0.30$& 0&...           &0&...           &0&...                     \\
10           &9  &$14.13\pm0.22$&0&...           & 0&...           &0&...           &0&...                     \\
11           &...&...           &0&...           & 0&...           &0&...           &0&...                     \\
12           &6  &$14.22\pm0.29$&0&...           & 0&...           &0&...           &0&...                     \\
13           &...&...           &0&...           & 0&...           &0&...           &0&...                     \\
14           &6  &$14.06\pm0.23$&0&...           & 0&...           &0&...           &0&...                     \\
15           &...&...           &0&...           & 0&...           &0&...           &0&...                     \\
16           &4  &$14.09\pm0.30$&0&...           & 0&...           &0&...           &0&...                     \\
17           &...&...           &0&...           & 0&...           &0&...           &0&...                     \\
18           &3  &$13.87\pm0.30$&0&...           & 0&...           &0&...           &0&...                     \\
19           &...&...           &0&...           & 0&...           &0&...           &0&...                     \\
20           &3  &$13.76\pm0.30$&0&...           & 0&...           &0&...           &0&...                     \\
21           &...&...           &0&...           & 0&...           &0&...           &0&...                     \\
22           &3  &$13.47\pm0.30$&0&...           & 0&...           &0&...           &0&...                     \\
23           &...&...           &0&...           & 0&...           &0&...           &0&...                     \\
24           &2  &$13.44\pm0.30$&0&...           & 0&...           &0&...           &0&...                     \\
$\log N_{\rm obs}^{d}$&71&$15.29\pm0.10$&7&$13.46\pm0.12$&52&$15.44\pm0.09$&7&$13.86\pm0.06$&5&$17.99\pm0.13$  \\
$\log N_{\rm int}^{e}$&  &$15.34\pm0.10$& &$13.79\pm0.12$&  &$15.44\pm0.06$& &$13.86\pm0.09$& &$18.12\pm0.13$  \\
$T_{\rm rot     }$ [K]&  &$328  \pm37  $& &$256  \pm30  $&  &$11.5 \pm0.2 $& &$ 8.0 \pm0.6 $& &$51   \pm37 $   \\
             &   &              & &              &  &              & &              & &                         \\
\hline
             &   &              & &              &  &              & &              & &                         \\
\multicolumn{4}{l}{C$_{2}$/C$^{13}$C$^{f}$     }&$ 36\pm 13$&&     & &              & &                         \\
\multicolumn{4}{l}{CN/$^{13}$CN$^{f}$          }&$ 38\pm  2$&&     & &              & &                         \\
\multicolumn{4}{l}{C/$^{13}$C$^{f}$            }&$ 72\pm 16$&&     & &              & &                         \\
\multicolumn{4}{l}{CO/(C$_{2}$+C$^{13}$C)$^{f}$}&$606\pm230$&&     & &              & &                         \\
\multicolumn{4}{l}{CO/(CN+$^{13}$CN)$^{f}$     }&$475\pm175$&&     & &              & &                         \\
             &   &              & &              &  &              & &              & &                         \\
\hline
\hline
\end{tabular} \\
$^{a}$ number of lines used in averaging. \\
$^{b}$ if less than 5 lines used, then error is 0.30~dex. \\
$^{c}$ from Paper~III. \\
$^{d}$ summed over all observed transitions [cm$^{-1}$]. \\
$^{e}$ integrated using the rotational diagram and an 
infinite number of transitions [cm$^{-1}$]. \\
$^{f}$ column density ratios derived from $\log N_{\rm int}$.
\end{table*}

\clearpage

\clearpage

\clearpage

\end{document}